\documentclass[twoside,12pt]{article}
\usepackage{epsfig,graphicx,amsmath,amssymb,url}

\def\Journal#1#2#3#4{{#1} {#2} (#4) #3 }

\def\PRD{{\em Phys. Rev.} D}

\newcommand{\be}{\begin{equation}}
\newcommand{\ee}{\end{equation}}
\newcommand{\bea}{\begin{eqnarray}}
\newcommand{\eea}{\end{eqnarray}}

\begin{document}

%\title{ \vspace{1cm} New Nuclei around the N = Z in the A = 80-90 region}
\title{ \vspace{1cm} Fast Radio Bursts}
\author{J.\ I.\ Katz,$^1$\\
\\
$^1$Department of Physics and McDonnell Center for the Space Sciences\\
Washington University, St. Louis, Mo. 63130 USA}
\maketitle
\begin{abstract} 
More than a decade after their discovery, astronomical Fast Radio Bursts
remain enigmatic.  They are known to occur at ``cosmological'' distances,
implying large energy and radiated power, extraordinarily high brightness
temperature and coherent emission.  Yet their source objects, the means by
which energy is released and their radiation processes remain unknown.  This
review is organized around these unanswered questions.
\end{abstract}
%\eject
%\tableofcontents
\section{Introduction}
The first Fast Radio Burst (FRB) was discovered in 2007 \cite{L07}.  They
were not universally accepted as a real astronomical phenomenon until
confirmed by the identification of five more FRBs several years
later \cite{KSKL12,T13}.  Most of the early discoveries were made in
archival data from the Parkes Multibeam Pulsar Survey, whose 13 beams survey
the sky for unknown sources 13 times faster than would a single beam.  Even
today, 25 of the 34 FRBs in the FRB Catalogue \cite{FRBCat} were discovered
at Parkes.  Slow acceptance was the result of the well known problem of
electromagnetic interference, mostly anthropogenic but also including
natural phenomena like lightning, in transient radio-frequency observations.
In fact, the anthropogenic ``perytons'' \cite{BS11}, not understood in
detail but later demonstrated to be produced by microwave ovens \cite{P15},
and bearing some resemblance to the first FRB discovered, were a source of
skepticism.

Since the general acceptance of FRBs, the literature, both observational and
theoretical, has exploded.  Searches on the SAO/NASA Astrophysics Data
System \cite{ADS} produce hundreds of citations, the number depending on how
the search criteria are defined and whether meeting abstracts and similar
unrefereed publications are included along with papers in archival journals.
It would not be useful to attempt to survey this entire literature; modern
bibliographic tools enable anyone to conduct such a search easily, and such
a survey would become obsolete in a few months.  Instead, I will try to
frame the major questions, the hypotheses offered to answer them, and
possible means of testing these hypotheses.  This involves subjective
judgment, and I apologize to those whose work I have neglected.

Astronomy is an observational science, but the ability to design an
observational program introduces an experimental aspect: carrying out such
a program and comparing its results to predictions is analogous to
performing a laboratory experiment.  The chief difference is that many
astronomical theories make, at best, qualitative predictions.  In astronomy
initial conditions are infrequently known, their effects persist through
the life of the system under study, and often essential processes are
``turbulent''.  ``Turbulence'' extends far beyond the homogeneous stationary
incompressible turbulence understood by Kolmogorov to include almost any
complex hydrodynamic or plasma process; it often means ``too complicated
and uncertain to calculate''.

The Solar neutrino problem is a striking exception: The Sun has lost memory
of its initial conditions, except for its mass and chemical composition, and
turbulence (in its convective zone) makes very little difference to its
properties.  No such luck applies to most astronomical phenomena, other
than stellar structure and celestial mechanics.  55 years after the
recognition of active galactic nuclei (``quasars'') and 50 years after the
discovery of radio pulsars, we have only the most qualitative understanding
of how they work and no consensus as to even the basics of pulsar
electrodynamics or why they emit observable pulses.  Active galactic nuclei
involve turbulent accretion flows and pulsars involve plasma turbulence and
coherent emission; we are now warned that such processes have been
particularly difficult to understand.  Phenomenology may be all that we can 
hope for.

Recent progress in data analysis makes this an opportune time to review
Fast Radio Bursts.  The UTMOST \cite{UTMOST} processor at Molonglo and the
Breakthrough Listen \cite{Breakthrough} processor at Green Bank have 
unprecedented spectral (100--200 kHz) and temporal ($10\,\mu$s) resolution
that resolve the frequency and temporal structure of bursts, revealing their
fine-scale dependence on both variables \cite{F18,G18}.  No longer are FRBs
described only by a single width of $\gtrsim 1\,$ms and spectral resolution
of multiple MHz, and this sharper temporal and spectral resolution has made
it possible to separate the effects of scintillation from the intrinsic
properties of the bursts.  Spectral and temporal complexity, first found
in FRB 121002 \cite{Ch16} and the repeating FRB 121102 \cite{S16}, now
appear to be universal characteristics of FRB.

The number of bits of information ideally obtainable from a burst
of flux $F_\nu(t)$ and fluence $\cal F$ over a spectral width $W_\nu$ and
temporal width $W_t$, observed with frequency resolution $\Delta \nu$ and
temporal resolution $\Delta t$ by an antenna of effective area $A$ and
system temperature $T_{sys}$ (typically about 25 K), is \cite{CT06}
\begin{equation}
	N_{bits} = {1 \over 2} \sum_{i,j} \log_2{\left(1 + {F_{\nu_i}(t_j)
	\Delta\nu\,\Delta t} {A \over k_B T_{sys}}\right)} \sim {1 \over 2}
	{W_\nu W_t \over \Delta\nu\,\Delta t}
	\log_2{\left(1 + {{\cal F} \Delta \nu \Delta t \over W_\nu W_t}
	{A \over k_B T_{sys}}\right)}.
\end{equation}
The argument of the logarithm is $1\ +$ the signal to noise ratio.  The
logarithm is summed over all independent channels, defined by widths in
frequency and time.  In the final approximate expression the sum is
approximated by multiplication by the number of independent channels.
The number of independent channels increases much more rapidly as $\Delta
\nu$ and $\Delta t$ decrease than the logarithm decreases, so that improving
resolution increases the information content of the signal.  This has
revealed the spectral and temporal complexity of FRBs.

The literature survey for this review was completed April 12, 2018.  I
apologize for the neglect of some papers that appeared prior to that date.
This review does not contain a complete bibliography, but rather a critical
assessment, and judgment as to how much observational detail or theoretical
speculation is appropriate is necessarily subjective.  I have been
willing to consider speculations about hydrodynamics and plasma physics,
that often behave in unpredictable and mysterious ways ({\it e.g.,\/}
accretion discs and coherent pulsar radiation), but not about particles or
objects for whose existence there is no empirical evidence.  The focus of
this review is on FRBs and their mechanisms, to the exclusion of their use
to study cosmology, the intergalactic medium or other problems.
\section{Basic Observational Facts}
\label{facts}
\subsection{Rate}
\label{rate}
It is hard to define the rate of FRB because there is no evidence that they,
unlike supernov\ae\ or mergers of black holes or neutron stars, are discrete
events with a well-defined rate.  Weaker FRB appear to vastly outnumber
stronger FRB \cite{K16b,K17d}, making any value of their rate dependent on
the detection threshold (as well as on frequency); a better and more popular
analogy is with stellar flares or outbursts.  An early rate estimate
\cite{T13} based on observations at Parkes was $\sim 3 \times 10^6$ y$^{-1}$
with a very large uncertainty, over the whole sky.  Observations with the
more sensitive Arecibo observatory \cite{S14} indicated a much higher rate
but with an uncertainty of at least an order of magnitude.  The much lower
rate of actual detections is accounted for by the fact that radio telescope
beams cover only a very small fraction of the sky at any time.

Taking $\sim 10^{11}$ galaxies in the observable Universe (at redshifts
$\lesssim 1$, as indicated for most FRB; Sec.~\ref{distance}) then suggests
a FRB rate of $\sim 3 \times 10^{-5}$ galaxy$^{-1}$ y$^{-1}$.  FRB strong
enough to be detected must be rare events, rarer (if roughly isotropic
emitters) than supernov\ae.  They may be products of very unusual
circumstances.
\subsection{Dispersion}
\begin{figure}
	\centering
	\includegraphics[width=0.99\columnwidth]{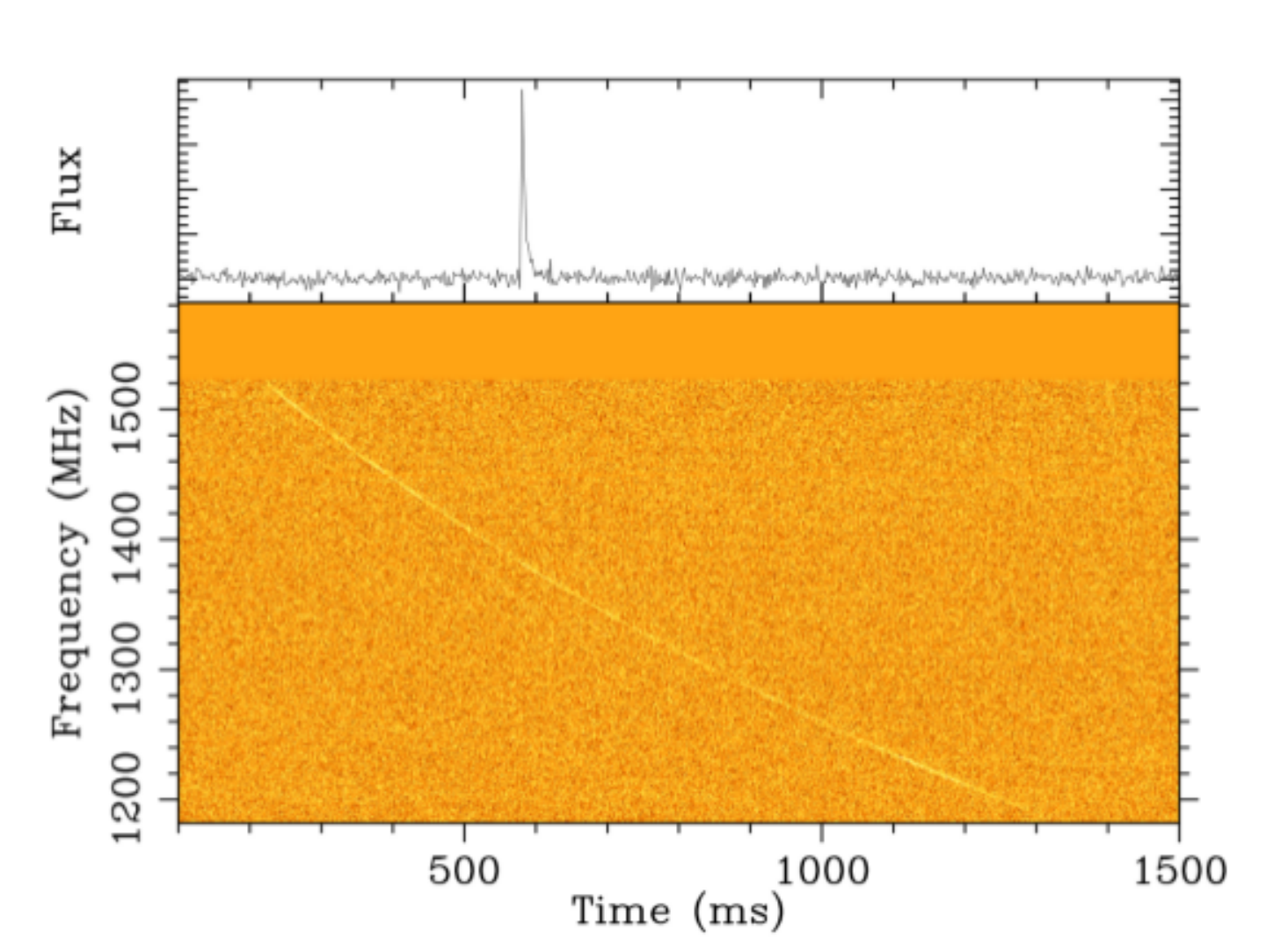}
	\caption{\label{110220} Upper panel is the frequency-integrated (over
	the 1182--1518 MHz bandpass) flux of FRB 110220 as a function of
	time after the dispersion measure has been fitted and the dispersion
	removed; time zero is arbitrary \cite{FRBCat}.  The pulse has a
	decaying ``tail'' as a result of multipath propagation in a
	turbulent medium; the pulse width depends on frequency \cite{T13}
	in accord with prediction.  As a result, the shape of the pulse
	depends on the bandpass over which it is integrated.  Lower panel is
	a frequency-time (``waterfall'') plot of spectral brightness
	$F_\nu(t)$ of FRB 110220, displayed as a gray scale, replotted from
	data of \cite{FRBCat}.  The general speckle is thermal noise in the
	detector system and frequencies above 1518 MHz were removed.  The
	curved trajectory in $(\nu,t)$ space results from dispersion
	(frequency-dependent group velocity) between the source and the
	observer.  There is structure in the received spectrum on scales
	$\sim 10\text{--}100$ MHz.}
\end{figure}

The most striking feature of a FRB, and one that immediately distinguishes
it from almost all electromagnetic interference, is the frequency-dependent
arrival time of its energy, as shown in Fig.~\ref{110220}.  This is familiar
to radio astronomers because it occurs when a radio-frequency signal
propagates through the interstellar plasma; it is a basic and readily and
accurately measured parameter of any rapidly varying radio source, such as a
pulsar or a FRB, and often the first parameter quoted.  The time delay of
the arrival of a signal of frequency $\nu$ with respect to the arrival of an
infinite-frequency signal, after passage through a low density plasma
($\omega \gg \omega_p$, where $\omega_p = 4 \pi n_e e^2/m_e$ is the plasma
frequency and $n_e$ the electron density) is
\begin{equation}
	\label{DM}
	\Delta t = {e^2 \over 2 \pi m_e c \nu^2} \int\!n_e(\ell)\,d\ell =
	{e^2 \over 2 \pi m_e c \nu^2} \text{DM}.
\end{equation}
The dispersion measure $\text{DM} = \int\!n_e(\ell)\,d\ell$ is usually given
in the astronomically convenient units of pc cm$^{-3} = 3.086 \times
10^{18}$ cm$^{-2}$.  At higher densities ($n_e \gtrsim 10^7$ cm$^{-3}$ for
GHz radiation, not encountered in the interstellar or intergalactic medium)
Eq.~\ref{DM} must be replaced by an expansion in powers of
$(\omega_p/\omega)^2 \ll 1$; the fact that the observed exponent of $\nu$ is
$-2$ to an accuracy of about $\pm 0.005$ sets upper bounds on the density of
the plasma through which the radiation traveled.

The first FRB to be discovered \cite{L07} aroused initial skepticism, in
part because of the well-known problem in radio astronomy of electromagnetic
interference, in part because it is natural to be skeptical of a single
observation until it is reproduced, and in part because it was much brighter
than the detection threshold---where were its fainter analogues that would
be expected to be more numerous, just as faint stars are more numerous than
brighter stars?  Skepticism was enhanced by the discovery of the
anthropogenic perytons \cite{BS11} that appeared to be dispersed with
qualitatively the same dispersion as the FRB.  The question was not whether
perytons were astronomical phenomena (it was evident that they were
interference), but whether the FRB was a peryton.  The discovery of several
more FRBs with a wide range of dispersion measures \cite{KSKL12,T13}, as
well as the observation that the pulse shape of the first FRB showed
evidence of scattering along its propagation path \cite{L07} dispelled any
doubt that FRB are real astronomical phenomena.

The dispersion measures of known FRBs are in the range 177--2596 pc
cm$^{-3}$ \cite{FRBCat}.  Some
portion of this, typically less than 40 pc cm$^{-3}$ but much greater if the
FRB was observed through a long path in the plane of our Galaxy, is
attributable to Galactic plasma.  This plasma has been well studied because
it disperses pulsar signals \cite{CL02}.  The remainder includes an unknown
contribution from near the source and a contribution from the intergalactic
plasma.  If the near-source contribution (that may include matter
distributed throughout the host galaxy or surrounding clouds, so that it
need only be more concentrated than the mean intergalactic plasma) is small,
the remainder may be attributed to intergalactic plasma.  This is usually
assumed, on the tacit assumption that the near-source region resembles
our Galaxy that would contribute comparatively little to the dispersion
measure, but this assumption may not always be justified \cite{C16}.
\subsection{Distances}
\label{distance}
\begin{figure}
	\centering
	\includegraphics[width=0.99\columnwidth]{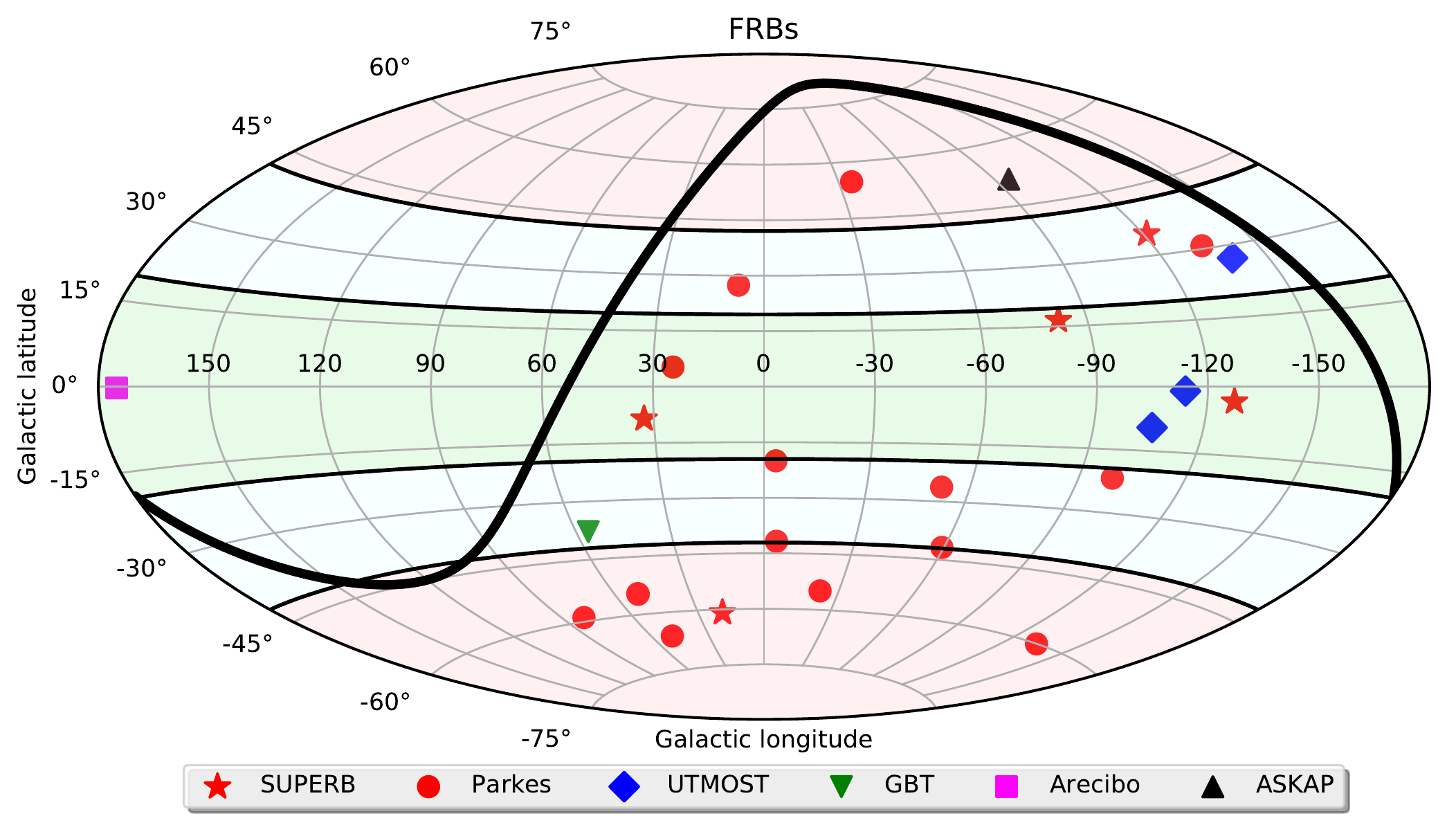}
	\caption{\label{bhandari} Distribution of FRBs discoveries on the
	sky in Galactic coordinates, replotted from data of \cite{B18}.  The
	thick black line is the horizon limit of the Parkes radio telescope
	in Australia, the source of most (SUPERB uses the Parkes telescope
	and UTMOST is at a similar latitude in Molonglo, Australia) of the
	detections.  Nearly all the discoveries were made in the Southern
	hemisphere, with the exceptions of one each from Arecibo and
	Green Bank (GBT).  FRB are not concentrated in the Galactic plane,
	excluding Galactic origin.  Some dependence on Galactic latitude
	has been reported \cite{P14}, and possibly explained as a propagation
	effect \cite{MJ15}, but the statistics may be affected by nonuniform
	search and may be consistent with statistical isotropy.}
\end{figure}

It is believed that FRBs originate at ``cosmological'' distances, meaning
redshifts $z = {\cal O}(1)$.  Several arguments support this conclusion:
\begin{itemize}
	\item No FRB has been identified with any cosmologically local
		object (peculiar star, pulsar, binary X-ray source,
		galaxy with $z \ll 1$, {\it etc.\/}).
	\item The dispersion measures of FRBs can be attributed to
		the intergalactic plasma if they are at redshifts
		${\cal O}(0.2\text{--}2)$.  Although there is no direct
		evidence against attributing the dispersion to a
		near-source region, the empirical upper bound on
		the exponent in the dispersion delay $\Delta t \propto
		\nu^\alpha$ implies $n_e \le (m_e \omega^2/6 \pi e^2)
	        \max{(-\alpha-2)} \approx 8 \times 10^7$ cm$^{-3}$
		\cite{FRBCat,K14b}.  The requirement that the dispersing
		plasma not absorb the FRB radiation by the inverse
		bremsstrahlung process sets an additional
		temperature-dependent upper bound on the plasma density in
		the region that contributes most of the dispersion measure:
		$n_e \lesssim 0.5 m_e^{3/2} c \nu^2 (k_B T)^{3/2}/(e^6 g
		\text{DM}) \approx 3 \times 10^4 \nu_{1400}^2 T_4^{3/2}/
		\text{DM}_{1000}$ cm$^{-3}$ where $\nu_{1400} = \nu/$1400
		MHz, $T_4 = T/10^4$\,K, $\text{DM}_{1000} = \text{DM/1000
		pc-cm}^{-3}$ and the Gaunt factor $g \approx 1.1$
		\cite{S62}.  These bounds exclude dense stellar coron\ae,
		winds or ejecta (such as supernova remnants) as the origin
		of the dispersion.
	\item Near-source clouds resembling known interstellar clouds would
		have insufficient DM, and hypothetical more massive or
		larger clouds would rapidly collapse under their own
		gravity \cite{K16b}.  While cosmology naturally leads to $z
		\sim 1$ and $\text{DM} \sim 1000$ pc cm$^{-3}$, typical of
		FRB, because the Universe evolves on that scale, no such
		natural scale of DM is apparent for a dense cosmologically
		local cloud.  In standard cosmology the intergalactic
		contribution is $\approx 1000\,z$ pc-cm$^{3}$ for $z \lesssim
		5$ \cite{I03,I04} so that with these assumptions $z$ may be
		inferred from the measured DM.
	\item FRB are not concentrated in the Galactic plane, in contrast to
		Galactic objects like pulsars (Fig.~\ref{bhandari};
		\cite{B18}).  Any possible anisotropy \cite{P14} may be
		accounted for by the effects of Galactic propagation
		\cite{BSB14,MJ15} with an underlying isotropic distribution.
		This is difficult to quantify because the total number of
		detected FRB is small, the sky was not searched uniformly
		and there may be other selection effects.  
	\item One FRB (121102) has been observed, over several years, to
		repeat.  This permitted accurate (to a small fraction of an
		arc-sec) interferometric localization and identification
		with a rapidly star-forming dwarf galaxy and with a
		persistent radio source \cite{C17,M17,T17,B17}.  This galaxy
		has a redshift $z = 0.193$, demonstrating a cosmological 
		distance (and a significant near-source contribution to its
		total DM of 559.7 pc cm$^{-3}$ \cite{Mi18}).  Non-repeating
		FRBs resemble the repeater in most respects other than
		repetition (and rotation measure, discussed later) 
		\cite{F18,G18,S16,Mi18}, so that Ockham's Razor suggests
		that they are members of the same class of object, differing
		quantitatively but not qualitatively.  Perhaps the location
		of FRB 121102 in or near a galactic nucleus and association
		with a persistent radio source account for its (so far)
		unique repetition and large rotation measure, but without an
		accurate position for any other FRB this remains
		speculative.
\end{itemize}
\subsection{Energetics} Assuming that FRB emission is isotropic, if its
distance is known its luminosity and radiated energy directly follow from
its measured flux and fluence (time integral of the flux).  At smaller
distances space is nearly Euclidean and the inverse square law applies,
while at larger distances $z \gtrsim 1$ the curvature of space-time must be
considered \cite{I03,I04}.  Formally, this takes the form of an inverse
square law with a luminosity distance.  Typical results for the brighter
FRBs, assuming that their dispersion measures, aside from a small Galactic
contribution, are produced by the intergalactic medium, are luminosities
${\cal O}(10^{43})$ ergs/s and burst energies ${\cal O}(10^{40})$ ergs
\cite{T13}.  Fainter FRBs may be as much as two orders of magnitude less
powerful and energetic, but if any weaker they would not be detectable with
present instruments.

These energies may seem impressive, but are very small compared to those of
gamma-ray bursts, that may radiate $\sim 10^{51}$ ergs at a power of
$\sim 10^{50}$ ergs/s, or supernov\ae\ that may radiate as much energy,
but with a power $\sim 10^{44}$ ergs/s.  Of course, the inferred FRB
energies and power are only those in the frequency bands (800 MHz--8 GHz)
in which they are observed, with most observations between 800 MHz and 1.5
GHz.  No FRB has been detected outside the radio frequency range.  The
extraordinary sensitivity of radio telescopes, resulting from their large
collecting areas (as much as $\sim 10^5$ m$^2$) and absence of quantum
noise, imply that FRBs are much less energetic than other astrophysical
transients, and suggest that their counterparts in less favored parts of the
electromagnetic spectrum are unlikely to be detectable unless they are many
orders of magnitude more energetic than the FRBs themselves.

These energy and power estimates assume that FRBs radiate roughly
isotropically.  There is no evidence for this, and they should be properly
called isotropic-equivalent energy and power.  Of course, the mean FRB
energy flux in the present Universe is an empirical quantity, so that if the
true energy and power of FRBs are less because they are collimated emitters,
their number (or event rate) must be greater in inverse proportion to the
reduction in inferred energy and power.

It is remarkable that FRBs are the shortest known astronomical events, with
the sole exceptions of some pulsar pulses and their substructure
\cite{S04,HE07,HEJ16}, black hole and neutron star coalescences, and the
rise times (but not the durations) of Soft Gamma Repeater (SGR) outbursts.
FRB durations range from $\sim 30\,\mu$s to $\sim 20$ ms
\cite{FRBCat,G18,Mi18} and some longer bursts contain substructure as sharp
as $\sim 30\,\mu$s \cite{F18,G18,Mi18}, although some FRB durations are
significantly lengthened by multipath propagation \cite{T13}.  In
comparison, the shortest gamma-ray bursts are about 30 ms long and don't
display sharper substructure \cite{F94,Q13}, while most are either $\sim
1$~s long (``short GRB'') or tens of s long (``long GRB'').
\subsection{Brightness}
Radio astronomers describe sources by their brightness temperature,
defined as the temperature $T_b$ of a Wien Law emitter that would produce
the observed flux density $F_\nu$ (whose units are erg/cm$^2$-s-Hz):
\begin{equation}
	\label{brightness}
	k_B T_b = {1 \over 2} {F_\nu c^2 \over \nu^2}
	\left({D \over \Delta x}\right)^2,
\end{equation}
where the factor of 1/2 comes from assuming an unpolarized source and taking
$F_\nu$ to be the sum over polarizations, $D$ is the distance and $\Delta x$
is the source size.  $T_b$ is independent of $D$ (by Liouville's Theorem);
the explicit dependence on $D$ cancels the distance dependence of $F_\nu$.

To evaluate $T_b$ it is necessary to know the solid angle $(\Delta x/D)^2$
subtended by the source.  This cannot be measured directly, of course.  The
distance can be estimated by a variety of astronomical arguments (for FRBs,
by assuming that their dispersion is attributable to the intergalactic
plasma).  The source size is usually bounded $\Delta x \le c \Delta t$,
where $\Delta t$ is a pulse or sub-pulse width, though it is larger for
a source relativistically expanding toward the observer.  Nonthermal
astronomical sources are usually observed expanding toward the observer
because of the relativistic beaming of their emitted radiation.

The result for FRBs is $T_b \sim 10^{35}\,$K.  Of course, this does not mean
that any component is that hot, or even that any radiating particle has an
energy $\sim k_B T_b$; it only implies that the source is a coherent
nonthermal radiator, in which energy is radiated by charge ``clumps'',
regions of correlated charge.  This is not unprecedented; pulsars typically
have $T_b \sim 10^{26}\,$K and the nanoshots of some pulsars have $T_b \sim
10^{37}\,$K, greater even than that of FRBs.  High brightness temperatures
are even familiar: A radio station emitting a power $P$ into a bandwidth
$\Delta \nu$ from a half-wave antenna (roughly isotropically) has $T_b \sim
P/(k_B \Delta \nu) \sim 10^{24}\,$K if $P = 50$ kW and $\Delta \nu = 10$ kHz
(sufficient for voice and music).
\subsection{Spectra}
The spectrum $F_\nu$ of a radiation source conveys much information about
its astronomical nature and physical processes.  There is no evidence for
line spectra in FRBs, nor would that be expected because line radiation is
generally a thermal process with low $T_b$, although maser amplification can
occur (there are molecular masers in the interstellar medium).
Typically,\footnote{This is the definition of $\alpha$ used by \cite{S16},
but the opposite sign with $F_\nu \propto \nu^{-\alpha}$ is widely used in
the radio astronomical literature.} spectra of nonthermal sources are fitted
to a power law $F_\nu \propto \nu^\alpha$.  The spectra of incoherent
emitters of synchrotron radiation are usually well fit by power laws over
broad frequency ranges, and the slope $\alpha$ determines the energy
distribution of the radiating particles.  Breaks in the power law determine
breaks in the particle energy distribution that can be related to
characteristic acceleration and radiative energy loss times of the
particles, and hence to parameters (such as magnetic field and the level of
plasma turbulence) of the source region.

FRBs have been observed at frequencies from approximately 800 MHz
\cite{F18,M15} to 8 GHz \cite{G18}, but have not been detected at
frequencies below 100 MHz (by the Long Wavelength Array, LWA1) or at 15 GHz
\cite{L17} or between 100 and 200 MHz (by the Low Frequency Array
Radio Telescope, LOFAR \cite{Ka15}, or by the Murchison Wide Field Array,
MWA \cite{T15,Ro16}).  The absence of low frequency detections may be
attributed to multipath pulse broadening that is expected to increase
rapidly with decreasing frequency, typically with $\Delta t \propto
\nu^{-4}$ although pulsar data show a broad range of exponents
\cite{KMNJM15}.

FRB spectra have not been well described by power laws
\cite{L17}.  Values of $\alpha$ fitted to different bursts of the repeating
FRB 121102 have ranged from -10 to +14 \cite{S16}, far outside the range for
other radio sources, for which usually $-1 < \alpha < 0$, and far from
expectation for any physical process that produces a power law spectrum.
Large $\vert\alpha\vert$ is an indication of spectral structure on the
scale $\Delta \nu \sim \nu/\vert\alpha\vert$.

\begin{figure}
	\centering
	\includegraphics[width=0.84\columnwidth]{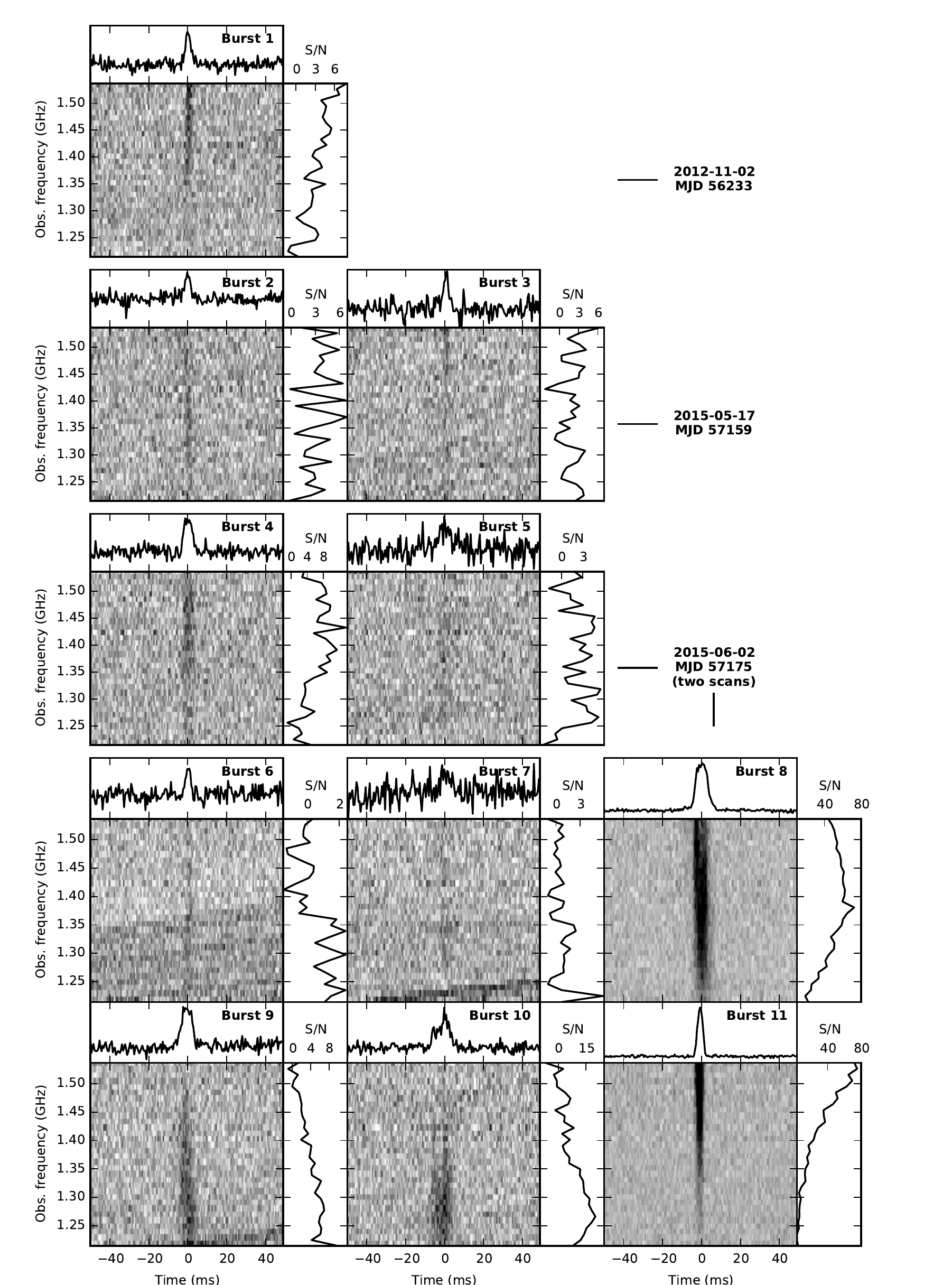}
	\caption{\label{spitlerf2} Frequency-time (``waterfall'') plots of
	bursts from the repeating FRB 121102.  Comparison of bursts 8--11
	shows that the spectra vary among bursts (this is best seen in the
	temporally integrated spectra plotted as $S/N$; comparison of the
	spectra of some bursts in the waterfall plots is obscured by
	gray-scale saturation).  Electromagnetic interference confined to
	sharply delimited frequency bands is evident in bursts 6, 7 and 9.
	In the plots it appears later at higher frequencies as a result of
	compensation for the fitted dispersion delay.
	P. Scholz private communication; replotted from data of \cite{S16}.}
\end{figure}

This has not been much remarked, although spectral structure has been
observed in almost every FRB (one exception was in the discovery paper
\cite{L07}, where it could not be seen because saturated one-bit data were
plotted).  For example, spectral structure is evident in Fig.~\ref{110220}
as a variation of brightness along the frequency-time curve.  Explicit plots
of time-integrated (to increase the signal to noise ratio) $F_\nu$ were
infrequently shown, although they are found more often in more recent
papers.  Fig.~\ref{spitlerf2} (the repeater FRB 121102) and Fig.~1 of
\cite{R16} (FRB 150807) very clearly show, in the brighter bursts with high
ratios of signal to noise, spectral structure on frequency scales $\sim
30\text{--}100$ MHz in the 1200--1500 MHz band.  This structure differs
among bursts of FRB 121102, all observed in the center of the beam pattern
(because the source's position is known the beam is steered to it) of the
same instrument \cite{S16}, confirming that the spectral structure is not an
instrumental artefact.  It has remained to distinguish intrinsic structure
of the emitted spectrum from chromatic scintillation, the result of the
fact that the plasma refractive index, and therefore the strength of a
signal received through a scattering medium, depend on frequency.
\subsection{Identifications with Other Objects}
Progress in interpreting astronomical discoveries usually depends on
identifying the newly discovered phenomenon with some other class of objects
about which more is known.  For example, radio sources were identified with
galaxies that have jets, double-lobed radio emitting regions and active
nuclei (``quasars'') \cite{N13}.  This led to the hypothesis \cite{S64},
confirmed decades later, that they are powered by accretion onto
supermassive black holes.  Pulsars were quickly identified with supernova
remnants \cite{P68}, confirming their identification as neutron stars and
explaining their origin.  More recently, the gravitational wave event
GW170817 was identified with signals across the electromagnetic spectrum,
including a visible ``kilonova'', its radio counterpart and a short
gamma-ray burst \cite{Br18}.

FRB 150418 has been associated with a fading radio transient in a galaxy at
redshift $z = 0.492$ \cite{Ke16}, a redshift that is consistent with the
measured dispersion measure of the FRB, although there is significant
uncertainty in the Galactic contribution and consistency requires that any
near-source contribution not be large.  Because this FRB has not been
observed to repeat it has not been possible to refine its position, whose
accuracy is described by the Parkes FWHM beam diameter of 14 arc-min.  The
reality of this association, although nominally statistically significant,
has not been widely accepted.

The only generally accepted localization of a FRB with another astronomical
object is that of the repeating FRB 121102 with a rapidly star-forming dwarf
galaxy at a redshift of $z = 0.193$ and with a persistent radio source,
likely in that galaxy, at a projected (transverse to the line of sight)
separation $< 40$ pc \cite{M17}.  It is unknown if there is a causal
relation between the FRB and the persistent source: if the persistent source
is a low-luminosity active galactic nucleus it may provide (by attracting
interstellar gas) an environment contributing to the activity of FRB 121102
without being directly involved; alternatively, the persistent source might
be a supernova remnant energized by a neutron star that also makes the FRB
outbursts \cite{MKM16,DWY17,KM17,MBM17}; in yet another hypothesis, an
intermediate mass ($10^2\text{--}10^6$ Solar mass) black hole both powers
the persistent source and is itself, with its accretion flow, the origin of
the FRB outbursts \cite{K17a}.

It is not known if those FRBs not observed to repeat are associated with
persistent radio sources because their positions on the sky are poorly
determined.  Obtaining accurate interferometric coordinates of FRB 121102
\cite{C17} required approximate coordinates (from earlier bursts) to direct
interferometry to the burst source.
\subsection{FRB Environments}
\label{environment}
\subsubsection{Non-repeating FRBs}
Little is known directly about the environments of most FRBs.  Many appear
to be broadened by multipath propagation (inferred from the roughly
$\propto \nu^{-4}$ frequency dependence of their widths \cite{T13}).  This
broadening is much greater than that of Galactic pulsars (except for the few
distant pulsars with very high DM lying in the Galactic plane) extrapolated
to the frequencies of FRB observations.  This argues that the broadening is
not attributable to our Galaxy, although there is a large scatter in the
broadening {\it vs.\/} DM relation of Galactic pulsars \cite{KMNJM15,B04}.
The FRB broadening does not depend monotonically on their dispersion measure
\cite{K16b}, which argues against an intergalactic origin if the DM is
largely intergalactic (however, see \cite{C16}) because the cosmological
uniformity of the intergalactic medium would predict such a monotonic
dependence.  This suggests that the scattering occurs in the near-source
region.  This region then must be significantly more turbulent (in the sense
of fluctuations of electron density that diffract and refract radio waves,
the ratio of broadening to DM) than the interstellar medium of our Galaxy,
suggesting a possible association with star-forming regions or even galactic
nuclei. 
\subsubsection{FRB 121102}
The exception to the lack of knowledge about FRB environments is the
repeating FRB 121102.  Its relation to the neighboring persistent source is
unknown, but reminiscent of the close proximity (0.1 pc projected distance)
of PSR J1745-2900 from the supermassive ($4 \times 10^6$ Solar masses) black
hole radio source Sgr A$^*$ at the Galactic center \cite{R13}.

FRB 121102 and PSR J1745-2900 have something else in common: large and
varying rotation measure (RM) \cite{G18,Mi18,E13,D18}, This parameter
describes the Faraday rotation of the direction of polarization of linearly
polarized radiation of wavelength $\lambda$ on passing through a magnetized
plasma:
\begin{equation}
	\label{RM}
	\theta(\lambda) = \lambda^2 {e^3 \over 2 \pi m_e^2 c^4}
	\int\!n_e B_\parallel\,d\ell \equiv \lambda^2\,\text{RM},
\end{equation}
where the integral $\int\!n_e B_\parallel\,d\ell$ is along the line of
sight.  Comparing the RM to the dispersion measure DM (Eq.~\ref{DM}) permits
an estimate to be made of a mean parallel component $B_\parallel$ of the
magnetic field.  Because of the weighting by the electron density $n_e$,
this mean measures the field in the densest regions, likely near the source,
and is essentially unaffected by the value of field elsewhere.

Both FRB 121102 and PSR J1745-2900 have very large RM.  The conventional
units are rad m$^{-2}$, so that RM is given in terms of the directly
measured quantity, $\theta(\lambda)$, while DM is given in terms of the
inferred quantity, electron column density.  $\text{RM} \approx
90,000$ rad m$^{-2}$ for FRB 121102 \cite{G18,Mi18}\footnote{Note that
\cite{Mi18} transform RM to the source frame, assuming, very plausibly, that
it is produced in a near-source plasma.}, several hundred times the RM of
other FRBs.  The RM of PSR J1745-2900 is about 70,000 rad m$^{-2}$
\cite{E13,D18}, hundreds of times typical pulsar RM and attributable to its
location near the Galactic center, though the relation of the strongly
magnetized plasma to the supermassive black hole is unclear.

The RM of both these objects has varied by 5--10\% over a time $t$ of
several months or a few years \cite{G18,Mi18,E13,D18}.  This indicates that
much of their RM is produced in a small region $\sim 10^{15}$ cm in size
(assuming velocities $v \sim 300$ km/s) in which conditions change rapidly.
By the Virial Theorem, this corresponds to a mass $M \sim v^3 t/G \sim 10^4$
Solar masses, and suggests such an intermediate mass black hole associated
with FRB 121102.  PSR J1745-2900 is about 0.1 pc (projected distance) from
the more massive black hole Sgr A$^*$ \cite{R13}, roughly consistent with
the assumed $v$, but the short $t$ requires that the variation be produced in
a region much smaller than 0.1 pc around PSR J1745-2900.

The facts that the RM of FRB 121102 and PSR J1745-2900 change by $\sim 10$\%
while there have been no corresponding observed changes in their DM,
indicate that the regions contributing to the changing RM contain only a
small fraction of the column density of electrons.  It is then possible to
estimate the magnetic field, assuming that on part of the line of sight DM
and RM change by comparable fractions, while on the remainder of the line of
sight DM is constant and the contribution to RM is negligible (because both
field and density are low).  The former describes a turbulent near-source
region and the latter the general interstellar or intergalactic medium:
\begin{equation}
	\label{B}
	B_\parallel \gtrsim {2 \pi m_e^2 c^4 \over e^3}\ {1\,\text{pc}
	\over 3.086 \times 10^{18}\,\text{cm}}\
	{\Delta \text{RM} \over \sup{(\Delta \text{DM})}} =
	12\,\text{milliGauss}\ {\Delta \text{RM} \over \sup{(\Delta
	\text{DM})}},
\end{equation}
where $\sup{(\Delta\text{DM})}$ is the least upper limit that can be placed
on any change in DM (none is observed, and the least upper limit is roughly 3
pc cm$^{-3}$).  The dimensional first factor equals $3.80 \times
10^{16}$ Gauss and converts $\Delta$RM from rad cm$^{-2}$ to Gauss cm$^{-2}$,
and the final result uses $\Delta$RM in the astronomical units of rad
cm$^{-2}$ and DM in pc cm$^{-3}$.  Eq.~\ref{B} is only a lower bound on
$B_\parallel$ because only an upper bound on $\Delta$DM is known; $\Delta$RM
could be much less and $B_\parallel$ much greater.

For both FRB 121102 and SGR J1745-2900 the inferred values of $B_\parallel
\gtrsim\text{several milliGauss}$.  This is enormous by interstellar
standards (where typical fields are 3--10$\,\mu$Gauss), and indicates that
these objects are found in dense, strongly magnetized regions.  It is
believed that interstellar fields are produced by turbulent dynamos, and
that rough equipartition between turbulent and magnetic energy densities
exists, so that high magnetic energy densities imply high plasma densities.
This is unsurprising at $\sim 0.1$ pc from the accreting supermassive
black hole at the Galactic center.  The similar inferred fields suggest that
FRB 121102 may be closely and causally associated with its persistent
neighboring source.

Hypotheses include that the persistent source is a young supernova remnant or
pulsar wind nebula energized by a neutron star that makes the radio
outbursts \cite{G18,M17,Mi18,DWY17,Be17}, that the FRB is only the neighbor
of a massive black hole that is the persistent source \cite{M17,Mi18}
(proximity, by immersing the FRB in a dense turbulent region might
contribute to its activity), and that accretion onto such a black hole makes
the bursts as well as the persistent source \cite{K17a}.

FRB 121102 differs from other FRBs in two ways: repetition, and a large RM
that implies a highly magnetized environment.  Is there a relation between
these properties?  This is impossible to prove, but one can speculate that
interaction with (accretion from?) its environment triggers its bursts.
A dense environment is plausibly more strongly magnetized than a lower
density environment (for example, if magnetic energy density scales with
thermal or kinetic energy density), and suggests a higher accretion rate.
FRB 121102 might be qualitatively similar to other FRBs, as suggested by
Ockham's Razor, but quantitatively different as a result of its environment.
Analogous speculations have been made in regard to SGRs, that they differ
from other neutron stars in being accompanied by small solid bodies whose
accretion produces their recurrent outbursts \cite{KTU94}.  This does not
require that FRB and SGR be associated, although that hypothesis has been
advanced (Sec.~\ref{SGR}).
\section{Models}
Because of the paucity of astronomical data concerning FRBs other than the
radio bursts themselves, it is necessary to resort to theoretical
arguments.  These cannot prove the origin of FRBs, but can suggest testable
hypotheses.  

FRBs are brief and energetic.  Although weak compared to gamma-ray bursts,
their radiated power, assuming isotropic emission, of up to $\sim 10^{43}$
ergs/s is comparable (briefly) to that of entire radio galaxies, though not
to the most luminous members of that class.  The combination of high power
and short time scales naturally points to association with neutron stars
because of their deep gravitational potentials, great gravitational,
magnetostatic and electromagnetic (radiation) energy densities and short
characteristic time scales ${\cal O} (R^3/GM)^{1/2}\sim 10\,\mu$s.  Many
possible models have been proposed that could have the energy, short time
scale and event rate of FRBs.

None of these models has been developed in sufficient detail to show that
it actually would make FRBs.  The obstacle is that coherent emission
requires plasma turbulence, and both the turbulence itself and the initial
conditions that create it are not understood.  The models have not gone
beyond {\em possibility}---showing that they cannot be disproved---to
demonstrate how and why they {\it should\/} make FRBs.

Because of these limitations of theoretical modeling, the comparison of
models to the observed properties of FRBs (Sec.~\ref{facts}) must be
phenomenological and will likely be qualitative.  Models that assume a
particular class of astronomical objects must be consistent with their known
properties and behavior.  The model must imply a rate of FRBs consistent
with their observed rate (Sec.~\ref{rate}).  

Such an object may become manifest in some
other manner.  For example, a rotating neutron star may imprint its
rotation period on its emission, or it may be the product of a supernova
that left a detectable remnant.  Proximity to an active galactic nucleus
(or other black hole with a radiating accretion flow) may be tested by
precise astrometry.  Young fast pulsars produce pulsar wind nebul\ae\ that
are steady radio sources, and soft gamma repeaters undergo infrequent but
very luminous outbursts.
\subsection{Pulsar Super-Pulses}
The brevity and high brightness temperature (Eq.~\ref{brightness}) of FRBs
point to an analogy with pulsars, suggesting that FRBs are super-pulsars of
some sort \cite{KSKL12,CW16,CSP16,LBP16}.  The existence of nulling pulsars
(pulsars with intervals during which no pulses are detected \cite{GJK12})
and rotating radio transients (RRAT) with rare pulses \cite{M09} suggests
that FRBs might be their extreme limit, with duty factors $\text{DF}
\lesssim 10^{-8}$ for non-repeating FRBs and $\text{DF} \sim
10^{-4}\text{--}10^{-8}$ for the repeating FRB 121102.

The fundamental assumption (essentially, the definition) of a pulsar model
is that the radiated power is drawn from the instantaneous spindown power
of a rotating neutron star.  There is, of course, magnetostatic energy, but
in pulsar models it is not a source of radiated energy.  As a result, pulsar
models must overcome a formidable energetic obstacle: their mean efficiency
of production of FRB power cannot exceed DF, and is likely much less.  The
efficiency of conversion of rotational energy to radio radiation in known
Galactic pulsars is $< 10^{-2}$, and is often orders of magnitude less
\cite{psrcat}.

\begin{figure}
	\centering
	\includegraphics[width=0.89\columnwidth]{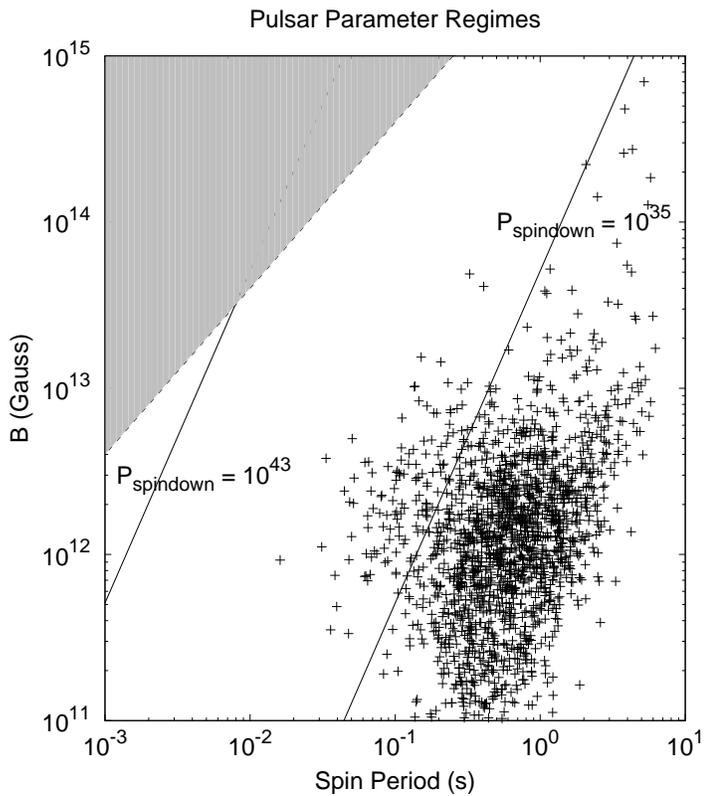}
	\caption{\label{PSR}Parameter regimes allowed for a pulsar model
	of FRBs.  Diagonal lines are labeled by the spindown power, an
	upper bound on the power that can be radiated as a FRB.  $+$
	indicate the parameters of Galactic pulsars.  The shaded region on
	the upper left is excluded by the requirement, applicable to the
	repeating FRB 121102 but not to non-repeating FRBs, that the spindown
	time not be less than five years, lower limit to any decay in its
	activity because it has been observed for five years.  Replotted
	from \cite{K17b}.}
\end{figure}

The instantaneous radiated power cannot exceed the pulsar spindown power
\begin{equation}
	P_{spindown} = {2 \over 3}{\mu^2 \Omega^4 \cos^2{\phi} \over c^3},
\end{equation}
where $\mu$ is the magnitude of the magnetic dipole moment, $\Omega$ is the
angular frequency of rotation, and $\phi$ is the angle between the dipole
and rotation axes.  This must supply the emitted FRB power that is as large
as $10^{43}$ ergs/s, divided by the unknown efficiency of radiation.  The
consequences are shown in Fig.~\ref{PSR}.

Although the na\"{\i}ve pulsar model can meet the energy requirements for
non-repeating FRBs, and, with optimistic assumptions as to efficiency, for
the repeating FRB 121102, it stretches the bounds of plausibility, and
demands extreme values for the pulsar parameters \cite{K17b,Ly17}.  No
neutron stars with such very fast rotation and high magnetic fields have
ever been observed.  That argument may not be compelling because such
pulsars would not spin fast for long, and might be very rare.  They have
been suggested to power supernov\ae\ \cite{OG71}, but the extreme values of
the parameters required to explain FRBs would imply deposition in the
supernova remnant of $\sim 10^{52}$ ergs of rotational energy, inconsistent
with observations of Galactic remnants.

The na\"{\i}ve pulsar model makes two critical assumptions:
\begin{itemize}
	\item There is no energy store that can be tapped; the radiated
		power is limited to the instantaneous spindown power.  This
		is consistent with observations of Galactic pulsars, but
		remains an assumption.
	\item Coherent radiation is emitted roughly isotropically. 
\end{itemize}
Relaxing these assumptions opens the possible parameter space and much
longer lifetimes are possible.  Fig.~\ref{PSR} shows the possible parameter
ranges if emission is strongly collimated so that $P \ll 10^{43}$ ergs/s.
Alternatively, if there is a store of energy that can be tapped on the time
scale of a FRB then these constraints are inapplicable.  For example,
transitions between different states of a neutron star magnetosphere might
release electrostatic energy, in an analogy to lightning \cite{K17c}.
\subsection{Soft Gamma Repeaters}
\label{SGR}
The extreme example of release of stored energy is the ``magnetar'' model
of Soft Gamma Repeaters \cite{K82,DT92,DT95}, in which the energy source is
the magnetostatic energy of a neutron star magnetosphere.
These models are essentially the opposite of pulsar models: The rotational
energy is negligible, and the radiated energy is drawn from the neutron
star's magnetostatic energy.  There is no fundamental limitation on the
radiated power, but the total magnetostatic energy cannot be released in a
time shorter than the light-crossing time of a neutron star magnetosphere
$\sim 30\,\mu$s. 

Many authors have suggested this as the origin of FRBs
\cite{MBM17,PP07,PP13,Ku14,L14,PC15,K16a,WY17}.  Energetics are not a
problem; SGR outburst energies have been as large as $\sim 10^{47}$ ergs,
seven orders of magnitude greater than those of FRBs.  A FRB might be an
epiphenomenon of a SGR, produced extremely inefficiently.

An additional argument for considering SGR-based models of FRBs is that
although SGR pulses are 0.1--0.2 s long, 10--100 times longer than FRBs, SGR 
rise very rapidly.  The rise time of the March 5, 1979 outburst of
SGR~$0525-66$ was $< 200\,\mu$s \cite{C80}, \cite{P05} reported an
exponential rise time of 300$\,\mu$s for the giant December~27, 2004
outburst of SGR~$1806-20$, while their published data suggest a value of
200$\,\mu$s, and the giant August~27, 1998 flare of SGR~1900+14 had a rise
time of $< 4\,$ms \cite{H99} and earlier outbursts had rise times
$\le 8\,$ms \cite{M79}.  These short rise times suggest a possible
connection.

Although the SGR hypothesis is attractive in some respects, there are 
objections to it:
%Rise time
\begin{itemize}
	\item SGR have thermal spectra peaking at $h \nu \sim 200$ keV,
		with no evidence for any nonthermal processes in the
		outbursts themselves (although the radio emission observed
		long afterward requires the acceleration of energetic
		particles).  However, the possibility that their sub-ms
		initial rise might be dominated by nonthermal processes
		with coherent emission cannot be excluded empirically
		because only $\sim 10^{-3}$ of their emission occurs during
		that rise and its contribution to the integrated spectrum
		may be undetectable.
	\item A source radiating at an intensity $\gtrsim 10^{29}$
		ergs/cm$^2$s is expected to form an equilibrium pair plasma
		by processes that turn two particles or photons into three
		(radiative Compton scattering, radiative pair production,
		three photon positron annihilation,
		{\it etc.\/}) \cite{K96}.  SGR during outburst exceed this
		intensity by orders of magnitude, arguing against them as
		the location of the coherently radiating nonequilibrium
		distribution of relativistic particles required for FRBs.
	\item The Parkes telescope was observing a pulsar at the time of the
		giant 27 December 2004 outburst of SGR 1806-20 \cite{TKP16}.
		The SGR was 31.5$^\circ$ above the horizon and 35.6$^\circ$
		away from the beam direction.  There was no evidence of a
		FRB, with an upper limit tens of dB lower than predicted for
		a Galactic FRB in the very far side lobes of the telescope's
		beam \cite{K16b,K14}.  Collimated emission by the FRB
		component of a SGR/FRB might be an explanation, but
		electrons accelerated and radiating upward would be
		accompanied by positrons accelerated and radiating downward,
		whose radiation would be broadly scattered by the neutron
		star or the dense SGR-radiating plasma.  Further, strong
		``magnetar'' fields would guide radiating electrons and
		spread their radiation pattern.
\end{itemize}
\subsection{Other Proposals}
Many other possible hypothetical FRB sources have been proposed.  Some are
listed here, roughly sorted into categories ordered by popularity (measured
by the number of papers).  Some straddle more than one category or don't
fall neatly into any category.  In general, only consistency with the
observed energetics, time scale and event rate has been demonstrated; the
hypotheses have not been proved wrong (those demonstrably wrong or
excessively speculative are not listed), but neither have they been proved
more than consistent with those constraints.
\subsubsection{Merging or colliding neutron stars}
The discovery \cite{HT75} of a binary system with a gravitational wave
lifetime shorter than the age of the universe and consisting of two neutron
stars was the motivation for construction of gravitational wave
observatories.  To date, one example (GW170817) of the predicted neutron
star mergers has been observed \cite{A17a}, and there will surely be more.
These mergers satisfy the criteria of brevity and sufficient energy to power
FRB, and have been suggested as their origin
\cite{To13,LP14,WYWDW16,DE17,YTK17}.  The observed neutron star merger rate
is several orders of magnitude less than the observed FRB rate
(Sec.~\ref{rate}), but these mergers can only be observed gravitationally at
much smaller distances than those estimated for FRB (Sec.~\ref{distance}).
There do not appear to have been any radio observations simultaneous with
GW170817, and follow-up observations found no FRB activity \cite{A17b}.
\subsubsection{Neutron star collapse}
A rotating neutron star that is above the upper mass limit of a non-rotating
neutron star will collapse as it loses angular momentum by radiating
magnetic dipole radiation (like a pulsar).  The collapse would be a sudden
event consistent with the brevity of a FRB, but could only occur once.  This
hypothesis might explain non-repeating FRBs but not the repeater.  Hence, it
would be unsatisfactory if the repeater is fundamentally the same object as
non-repeating FRBs, a plausible (because of their resemblance) but unproven
assumption.  It provides no evident mechanism of making FRBs because in such
a collapse the star is expected to ``wink out''---simply disappear as its
surface approaches the event horizon.  Temporal FRB substructure may be
particularly difficult to explain.  Several authors have advanced this
hypothesis \cite{FR14,Z14,RL14,DGS18,MNR18}.
\subsubsection{Interaction of a pulsar with its environment}
This hypothesis requires either that a pulsar wind interacting with its
environment be highly relativistic in order that the radiation received from
an extended region arrive in a time as short as a FRB \cite{DWY17,W17}, or
that an external influence act, over a short time, on a neutron star
magnetosphere \cite{Z17,Z18}.  These conditions may be met with suitable
values of the parameters, but the existence of these interactions and of the
flows required to produce them remain speculative.
\subsubsection{Binaries with neutron stars and other objects}
This category includes models in which a neutron star interacts with a
binary companion other than a neutron star.  It has been suggested in
several forms, in which the companion may be a white dwarf
\cite{GDLMW16,LCG18}, a black hole \cite{AB17,Bh17}, or an unspecified
variety of objects \cite{CYZ18}.
\subsubsection{Yet other neutron star models}
These hypotheses include the collision of asteroids with neutron stars
\cite{DWWH16,Ba17} (also suggested as the origin of SGR \cite{KTU94}).
Neutron starquakes \cite{WLYCLX18} and an analogy of lightning in a neutron
star magnetosphere \cite{K17c} have also been proposed.  Again, both the
existence of the proposed events and their ability to make FRBs are
speculative.
\subsubsection{Models without neutron stars}
Active galactic nuclei have been proposed as the origin of FRBs
\cite{VRBMV17}, as has an interaction between a white dwarf and a black hole
\cite{LHGL18}, an accretion funnel around an intermediate mass
($10^2\text{--}10^6$ Solar masses) black hole \cite{K17a}
(see Sec.~\ref{environment}) and yet other black hole models \cite{Z16}.
\section{Statistics}
Enough FRBs have been observed (33 at the time of writing \cite{FRBCat}), as
have been scores of outbursts of the repeating FRB 121102, to permit some
meaningful statistical studies.  
\subsection{All FRBs}
An earlier paper \cite{K16b} and review \cite{K16c} considered the
distributions of FRB widths, dispersion measures and fluence (more
accurately measured than flux in measurements of limited time resolution
and with significant detector noise).  The hypothesis that the dispersion
could be attributed to an expanding supernova remnant (SNR) was excluded:
if the SNR were young and compact enough to contribute significantly to the
dispersion, an excess of low-DM FRBs would be predicted, unless there were a
sharp cutoff on the age of FRB sources.  No such excess is
seen\footnote{This argument is entirely statistical, and independent of the
bounds placed on any expanding supernova remnant around the repeating FRB
121102.}.  Instead, the distribution of DM is consistent with a simple
cosmological model that attributes most of the dispersion to intergalactic
plasma, although a significant near-source contribution is not excluded.

The distribution of intensity or fluence $S$ is more controversial.  In a
homogeneous Euclidean universe (a fair approximation for redshifts $z
\lesssim 1$) the inverse square law predicts a relation between the
cumulative number $N$ brighter than $S$: $N \propto S^{-3/2}$.  This has
been tested, with some results finding consistency \cite{OCP16,ME18} but
others inferring an excess density of (anomalously bright) sources in the
local universe \cite{K17d}.  Further discussion of this distribution and of
the relation between DM and pulse broadening by scattering was provided by
\cite{R18}.
\subsection{The Repeating FRB 121102}
Scores of outbursts of FRB 121102 have been observed.  Its activity is
intermittent; during active periods bursts may be separated by tens of
seconds or a few minutes, while, with the same instrument, at other times
many hours may pass without a detected burst
\cite{G18,S16,Mi18,L17,Sc16,S17,H17}.  Several authors have shown that, as
is implied and evident from the preceding sentence, these data are
inconsistent with a stationary Poissonian process
\cite{WY17,WLYCLX18,CPO16,LWLBSB17,K18a}.  No theoretical model is well
enough developed for this to be regarded as either confirmation or
contradiction, but the statistics, including closely spaced (37 ms \cite{S17}
and 34 ms \cite{H17}) apparent pairs of bursts, may be consistent with a beam
executing a random walk in solid angle \cite{K17b,K18a}.
\section{Radiation}
The preceding discussions of FRB models did not address the mechanisms by
which these events may emit the radiation we observe.  A complete FRB model
would satisfy the requirements of burst energetics and number, but would
also identify the emission mechanism and explain why radiation with the
observed brightness and spectral characteristics is produced.
After 50 years of study, no pulsar model meets all these requirements, so it
is unrealistic to expect a FRB model to do better.  In practice, the demand
on models has only been that they are not demonstrably incapable of making
FRBs, and the literature is filled with models passing that relaxed test.

Consideration of any radiation mechanism must address two issues:
\begin{itemize}
	\item Are its intrinsic characteristics consistent with the
		observed spectra and polarization?
	\item Can it produce the observed power in plausible FRB
		circumstances?
\end{itemize}
The second issue has never been satisfactorily addressed because it depends
on the nonlinear evolution (saturation) of an unidentified plasma process.
In this section I discuss the first question, about which it is at least
possible to make some comments.  Neither issue has been completely resolved
for pulsars, which should make us humble about our ability to resolve them
for FRBs.
\subsection{Curvature Radiation}
A neutron star's magnetospheric energy density and available energy (in
either a pulsar mechanism or a SGR mechanism) are concentrated near the
star's surface; the magnetospheric energy density varies $\propto r^{-6}$,
the Poynting vector in the near zone varies $\propto r^{-5}$ and in the
radiation zone varies $\propto r^{-2}$.  The near-surface magnetic fields of
known neutron stars are in the range $10^7\text{--}10^{15}$ Gauss, so that
cyclotron or synchrotron frequencies, even of nonrelativistic particles, 
would be much too high to explain the GHz radiation of FRBs.  Hence modelers
almost universally appeal to the curvature radiation emitted by a particle
moving along magnetic field lines as the radiation process in FRBs
\cite{K17c,WYWDW16,GDLMW16,DWWH16,GL17,KLB17,YZ17,K18b,LK18}.  In order
to produce significant power emission must be coherent, with the fields of
many electrons (or positrons) adding in phase.

\begin{figure}
	\centering
	\includegraphics[width=0.99\columnwidth]{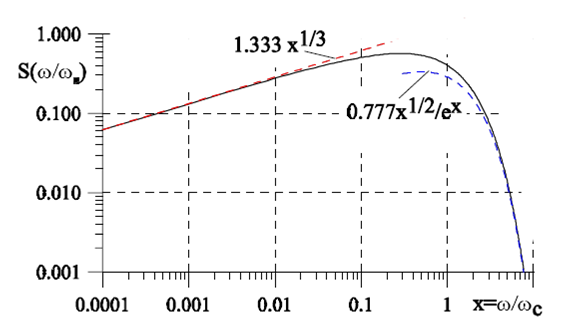}
	\caption{\label{curve} Spectrum of radiation emitted by a
	relativistic charged particle accelerated perpendicular to its
	velocity (synchrotron or curvature radiation) \cite{synchro}.  The
	characteristic angular frequency $\omega_c = 3 \gamma^3 c/(2\rho)$,
	where $\rho$ is the radius of curvature of its path.}
\end{figure}

A relativistic charged particle with Lorentz factor $\gamma \gg 1$ moving on
a path with radius of curvature $\rho$ radiates a broad spectrum, shown in
Fig.~\ref{curve}, peaked around the angular frequency $\omega_c =
3\gamma^3 c/(2\rho)$ \cite{J98}.  Aside from the different value of $\rho$
and the fact that in curvature radiation the particle's path is
instantaneously circular rather than (generally) helical, the properties of
synchrotron and curvature radiation are the same.  The radiation is strongly
linearly polarized, in agreement with most (but not all) FRBs\footnote{The
absence of linear polarization in some FRBs might perhaps be attributed to
an extremely high RM, but the required values would be orders of magnitude
larger than any known RM.}.

\begin{figure}
	\centering
	\includegraphics[width=0.99\columnwidth]{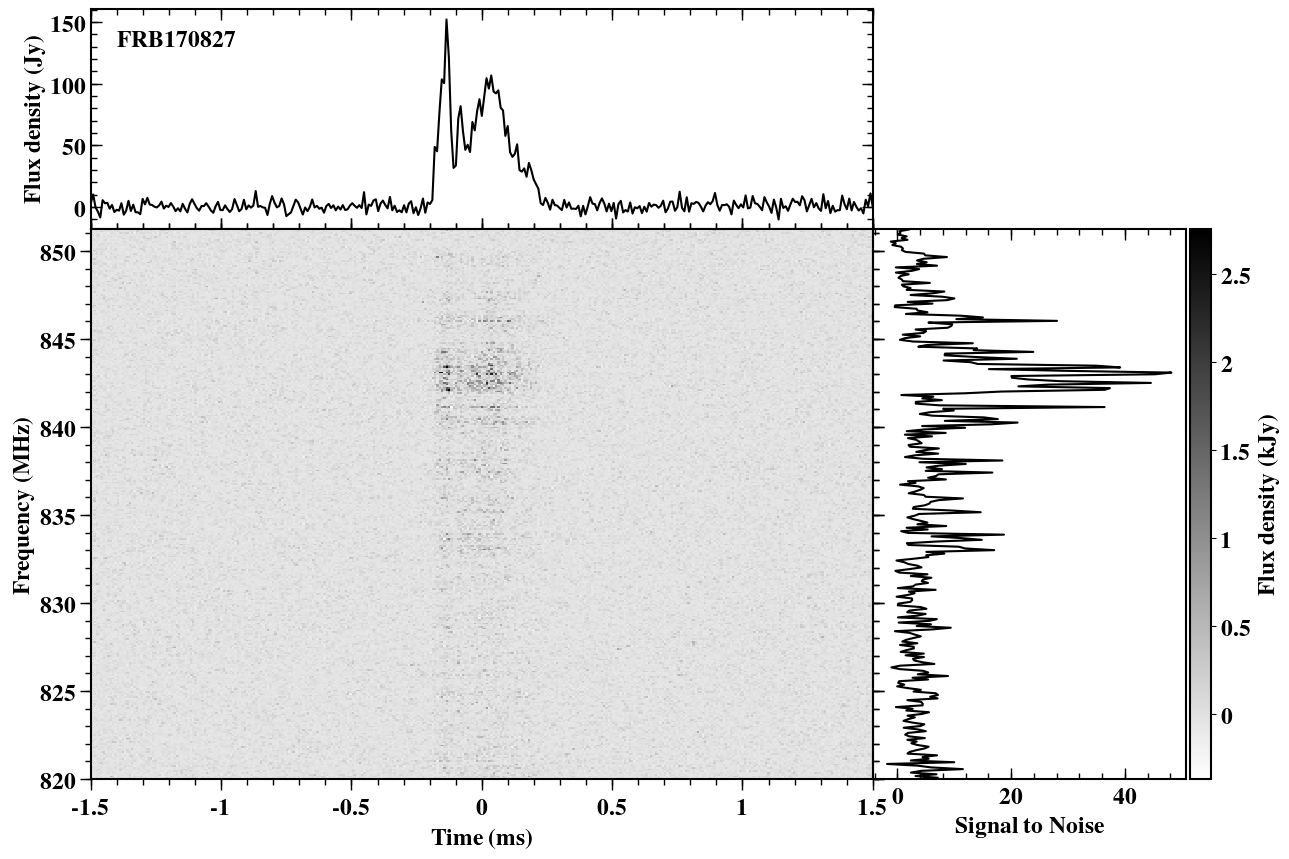}
	\caption{\label{170827} High resolution frequency-time
	(``waterfall'') plot of spectral flux density $F_\nu(t)$ of FRB
	170827, obtained by UTMOST; replotted from data of \cite{F18}.  The
	upper panel is the frequency-averaged flux density and the right
	hand panel is the time-integrated flux density, normalized by the
	system noise level.  Spectral structure on frequency scales from the
	resolution of 97.66 kHz to $\sim 1$--2 MHz is evident, as is
	temporal structure on time scales $\sim 10\,\mu$s.  The narrower
	structure of the spectrum persists through temporal dips in the
	intensity that extend across the entire (but only 31.25 MHz wide)
	spectral range of the observation.  This is consistent with the
	hypothesis that the narrower spectral structure is the result of
	scintillation produced by plasma diffraction along the line of
	sight.  In this small slice of the spectrum the broader structure
	(Figs.~\ref{110220},\ref{spitlerf2},\ref{57633.6}), that may reflect
	the intrinsic emission spectrum, is not evident.  Compare to FRB
	121102 (Fig.~\ref{gajjarf1}) whose spectrum (at much higher
	frequency) changes within a 2 ms burst, something difficult to
	explain by scintillation.}
\end{figure}

The extremely smooth and broad spectrum of curvature radiation, even of
monoenergetic particles, should be compared to
the spectral structure evident in Figs.~\ref{110220} and \ref{170827}.  The
observed data have structure that is inconsistent with the spectrum of
radiation emitted by charges moving along a circular (or helical, that
can be transformed to a frame in which it is circular) path.  Particles
moving on more complex paths, such as that of an undulator \cite{Song17},
produce spectrally structured radiation, but it is unclear how the complex
engineered magnetic structure of an undulator could occur naturally,
particularly in a neutron star magnetosphere where the field is dominated by
the intense static field produced by interior currents.
\subsection{Scintillation}
Radio radiation is refracted and diffracted passing through the
heterogeneous electron density of the interstellar medium, and this produces
scintillation, roughly analogous to the twinking of starlight \cite{R90}.
In contrast to refraction by air, that is only weakly dependent on
frequency, plasma refraction is strongly frequency-dependent.  Scintillation
appears in an instantaneous spectrum $F_\nu(t)$ or a time-integrated
spectrum $f_\nu$ as a dependence on $\nu$ that can be characterized by a
decorrelation bandwidth.  Scintillation is observed in the spectra and
time-dependence of radiation from Galactic pulsars that are often assumed to
radiate pulses of approximately constant energy (with the obvious exceptions
of nulling and giant pulses).  In FRBs much or all of the time-dependence is
intrinsic to the emitter, so that the observable effects of scintillation are
limited to the spectral dependence of the received energy.  Scintillation in
FRBs may be caused by near-source and intergalactic plasma as well as
Galactic plasma.

If the spectral structure of FRBs is attributed to scintillation, two 
predictions can be made:
\begin{enumerate}
	\item The spectrum will not change through a burst because the
		structure of diffracting and refracting plasma is unlikely
		to change significantly during a $\sim$ ms burst.  The
		spectrally-integrated flux varies because it is determined
		by the emitted power as well as by propagation effects.  As
		a result, the spectral and temporal distributions will be
		separable: $F_\nu(t) \propto f_\nu g(t).$
	\item The measured flux densities (as a function of frequency) will
		be distributed according to a Rayleigh distribution if
		scintillation is strong.
\end{enumerate}

The frequency and temporal dependences of $F_\nu(t)$ in a high
signal-to-noise burst of FRB 121102 (Fig.~\ref{gajjarf1}) show complex
structure.  Using a DM of 560 pc cm$^{-3}$, close to the value of 559.7
pc cm$^{-3}$ determined from a 30 $\mu$s burst \cite{Mi18}, the early
part of the burst radiates chiefly in a band 7000--7150 MHz, while about 2
ms later the radiation shifts downward in frequency and spreads to a band
5500--6600 MHz.  If all of the frequency structure is the result of
scintillation, the scattering screen must, implausibly, change its
configuration in $\sim 1$ ms.

\subsubsection{Narrow spectral structure}
\begin{figure}
	\centering
	\includegraphics[width=0.99\columnwidth]{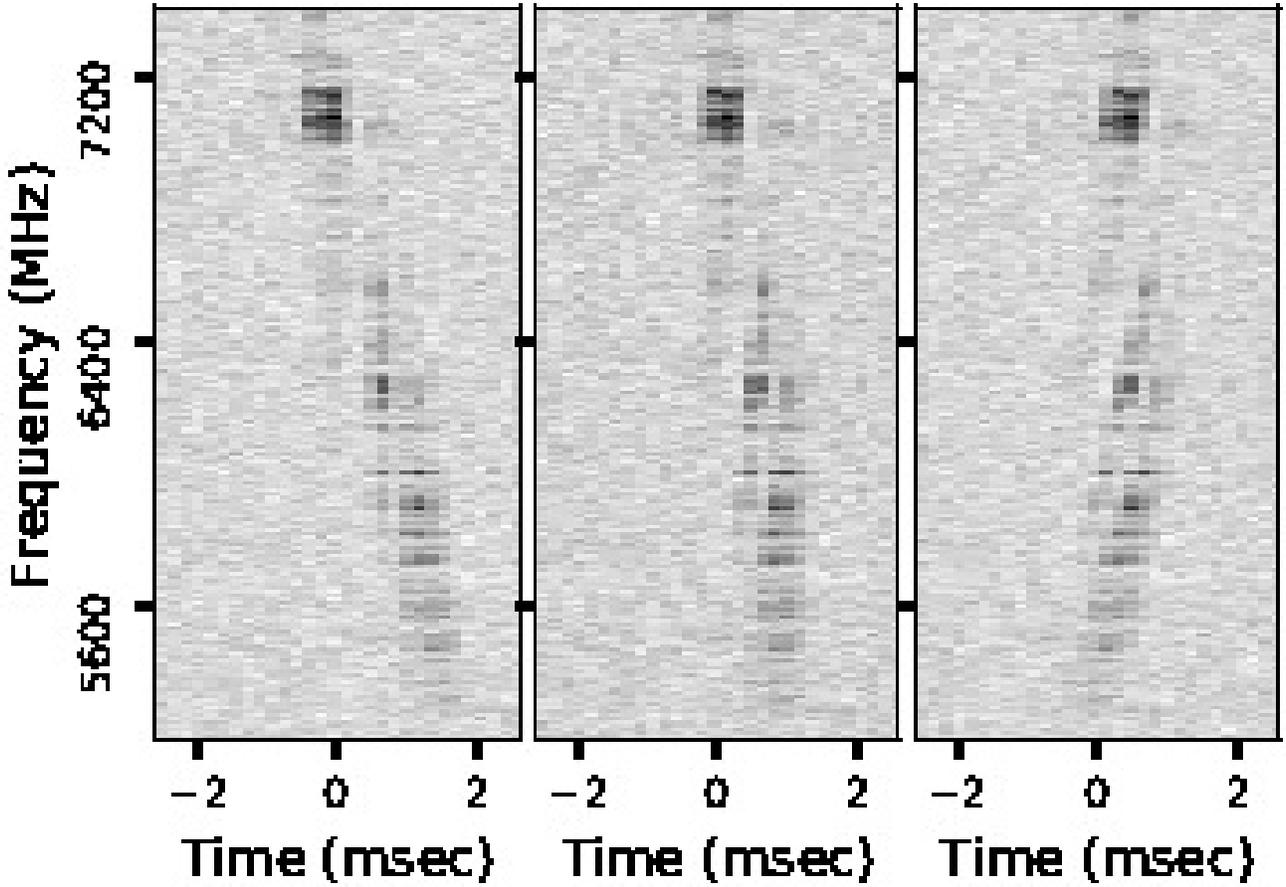}
	\caption{\label{gajjarf1} Time-frequency structure of a burst from
	FRB 121102, plotted for three assumed values (560, 580 and 600
	pc cm$^{-3}$, left to right) of DM (replotted from data of
	\cite{G18}; previous work found DM$= 559.7$ pc cm$^{-3}$
	\cite{Mi18}).  The broader spectral features change greatly within
	the 2 ms long burst, a challenge for scintillation models of
	spectral structure, but the narrower features do not.  The spectral
	structure also changes between bursts separated by times from tens
	of seconds to hours (Fig.~\ref{spitlerf2}, Figs.~1 and ED1 of
	\cite{Mi18} and Fig.~2 of \cite{G18}).  Compare to FRB 170827
	(Fig.~\ref{170827}) whose spectrum (at much lower frequency and in a
	narrower frequency range, perhaps precluding observation of broader
	features) changes less through the burst.  Alignment of the vertical
	stripes of zero intensity (assuming that flux nulls extend across
	the spectrum, as would be the case if they represent variations in
	output of the source) may permit accurate measurement of DM.
	Analogous but longer (minutes) nulls of emission by a radio pulsar
	are shown in Fig.~5d of \cite{R90}, where they are horizontal.}
\end{figure}

The known decorrelation bandwidths of FRB spectra are narrow.  For example,
in FRB 110523 it was $1.2 \pm 0.4$ MHz (at $\nu \approx 800$ MHz)
\cite{M15}, in FRB 150807 it was $100 \pm 50$ kHz (at $\nu \approx 1300$
MHz) \cite{R16} and showed the Rayleigh distribution of intensity predicted
for strong scintillation (Fig.~S9 of \cite{R16}), in FRB 170827 it was about
1.5 MHz (at $\nu \approx 830$ MHz) \cite{F18} and in pulses of FRB 121102 it
was $< 4$ MHz (at $\nu \approx 3$ GHz) \cite{L17}.  This narrow frequency
structure persisted through the varying intensity of a burst, appearing as
bright or dark horizontal lines in the high spectral and temporal resolution
``waterfall'' plots of Figs.~\ref{170827} and \ref{gajjarf1} and of
\cite{Mi18}.  This is naturally explained as the result of scintillation by
scattering screens at interstellar (or greater) distances from the source
and receiver that do not change on ms time scales.

\subsubsection{Broad spectral structure}
\begin{figure}
	\centering
	\includegraphics[width=0.99\columnwidth]{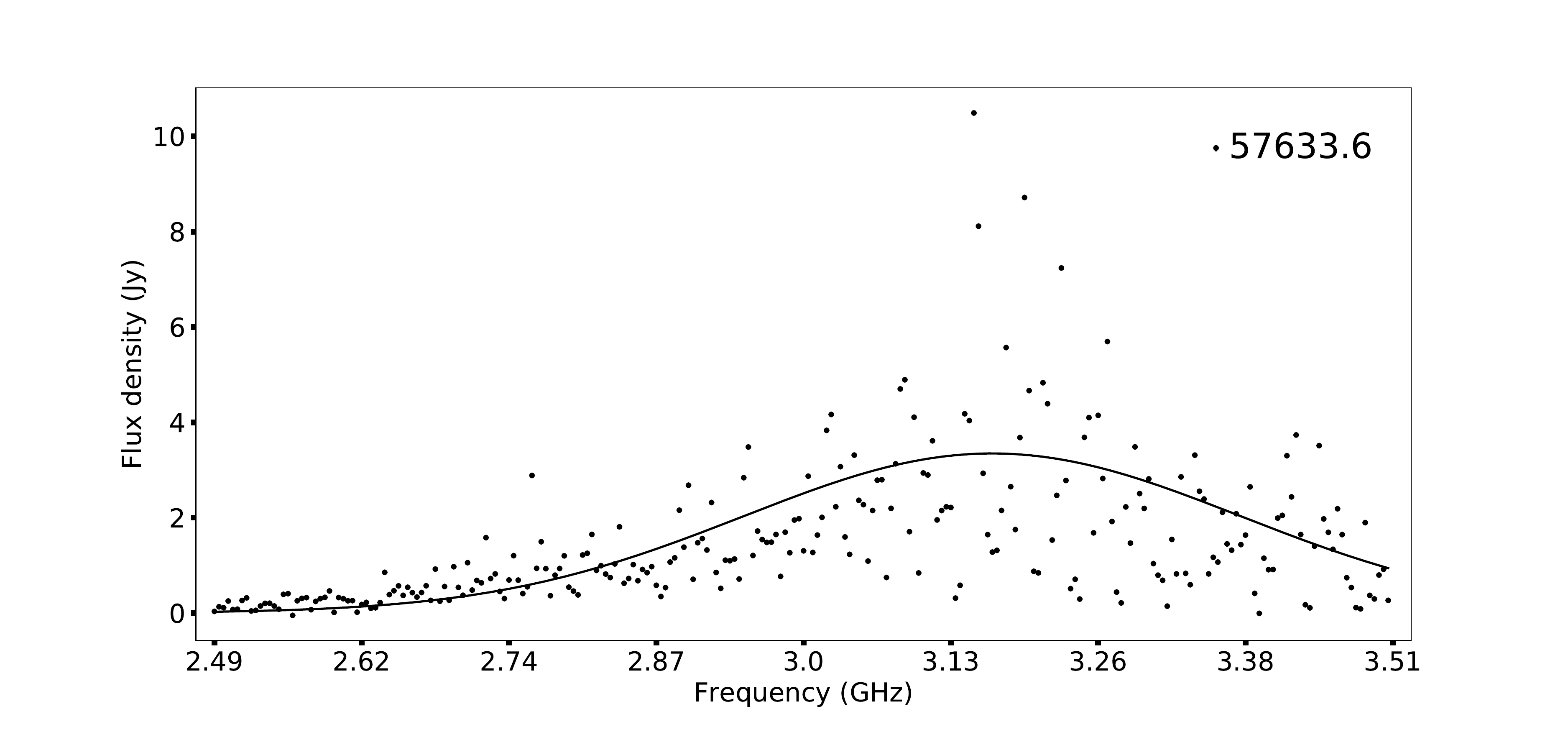}
	\caption{\label{57633.6} Spectrum of pulse 57633.6 of FRB
	121102 replotted from data of \cite{L17}.  The smooth curve is a
	fitted Gaussian.  The data are consistent with an emitted Gaussian
	function of frequency with half-width 250 MHz, multiplied by a
	Rayleigh distribution uncorrelated between 4 MHz wide frequency
	bins.  Scintillation with a decorrelation bandwidth much less than
	4 MHz would multiply the emitted spectrum by such a Rayleigh
	distribution.  The assumption of a Gaussian is arbitrary and other
	smooth functions with approximately the same width would also be
	consistent with the observations.  The noise level is indicated by
	the (tiny) error bar next to ``57633.6'' in the upper right.  Weaker
	pulses show similar behavior with different peak frequency and
	width, but with lower signal to noise ratio.}
\end{figure}

There is also spectral structure on broader scales of tens to hundreds of
MHz.  This is shown in Fig.~\ref{110220} for FRB 110220 \cite{T13,FRBCat},
Fig.~1 of \cite{R16} for FRB 150807, and Figs.~\ref{gajjarf1} and
\ref{57633.6} and \cite{Mi18} for bursts of FRB 121102 \cite{G18,L17}.
%  Two classes of
%explanation of this broader spectral structure can be considered.
%Sec.~\ref{plasma-curve} interprets it to be intrinsic to the radiation
%mechanism, disagreeing with the theoretical spectrum of radiation by an
%accelerated point charge shown in Fig.~\ref{curve}, but perhaps explicable
%as a result of a spatially structured radiating charge density.

A distant scattering screen producing a wide decorrelation bandwidth that
illuminates the screen that makes scintillation with a small decorrelation
bandwidth might produce broad spectral structure.  This hypothesis would
predict a Rayleigh distribution of the spectral power averaged over the wide
decorrelation bandwidth.  Such a distribution may be inconsistent with the
broad spectral regions of FRB 121102 with no detected flux (Fig.~4 of
\cite{L17}, Figs.~1c,d of \cite{Mi18} and Fig.~\ref{gajjarf1}).  A
quantitative test might be ambiguous because the spectral decorrelation
function of the scintillation would depend on uncertain assumptions about
the spatial structure of the scattering screen.  Also, the decorrelation
width of the broad scintillation would be a substantial fraction of the
observing bandwidth, so that only a few independent broad spectral bands
would be observed, making determination of the distribution of spectral
power problematic.

The broader frequency structure of FRB 121102 is difficult to explain as the
result of scintillation.  It varies greatly from burst to burst (Fig.~2 of
\cite{G18}) and even on sub-ms time scales within bursts
(Fig.~\ref{gajjarf1} and Figs.~1 and ED1 of \cite{Mi18}), with a
distribution of intensities unlike the Rayleigh statistics that are observed
(Fig.~S9 of \cite{R16}; Fig.~\ref{57633.6}) at frequency resolutions of a
few MHz or less.  In the nanoshots of the Crab pulsar the spectrum changes
on $\sim \mu$s time scales \cite{HE07,HEJ16}, an even more demanding
condition.  Such complex and rapidly changing spectra are very different
from the smooth spectrum of an accelerated point charge, and require
explanation.

%Alternatively, the temporal structure of FRB 170827 \cite{F18}, as fine as
%$\lesssim 10\,\mu$s and consistent across 31.25 MHz centered at 835 MHz,
%results from
%scintillation of a single source spectrum.  The dispersion measure of such a
%scattering screen must be $\lesssim 0.02\,\text{pc-cm}^{-3}$.  In addition,
%the intrinsic temporal width of the source must be smaller than the observed
%temporal width $\lesssim 10\,\mu$s of substructure, which is less than the
%neutron star light crossing time of $30\,\mu$s.  This is explained even if
%the entire magnetosphere is involved in emission if the radiating region
%(not only the radiating particles) moves from the neutron star surface
%towards the observer at relativistic speed, shortening the duration of the
%observed pulse.

\subsection{Coherent plasma-curvature radiation}
\label{plasma-curve}
Could the broader spectral features be intrinsic to the radiation mechanism,
disagreeing with the theoretical spectrum of radiation by an accelerated
point charge shown in Fig.~\ref{curve}, but perhaps explicable as a result
of a spatially structured radiating charge density?  If the scintillated
intrinsic spectrum of curvature radiation does not explain the frequency
structure of FRB bursts on scales of tens or hundreds of MHz, what does?

A possible answer may come from noting that the
brightness of FRBs requires coherent emission; charges must be clumped, with
net charge density greater than that resulting from an uncorrelated random
distribution of positive and negative charges (a thermodynamic equilibrium
distribution would have even smaller charge density fluctuations than those
of uncorrelated charges).  The origin of this clumping must be a plasma
instability, but its nature and properties are not known.  Very similar
considerations apply to pulsar radiation, as has been appreciated since
their discovery \cite{GZZ69}, and the spectral structure of pulsar nanoshots
\cite{S04,HE07,HEJ16} poses the same problems as the spectral structure of
FRBs.

Coherent radiation results from the acceleration of charge density
fluctuations (``clumps'') that contain so many charges that they are
described by a continuous function of space (the curved path along which the 
charges move) rather than by point charges.  Individual elementary charges
add a very low amplitude fine structure to this continuous distribution and
radiate incoherently, but at an undetectable level many orders of magnitude
less than the coherent radiation.  The radiation field is the convolution
of the field of radiation by a point charge with the spatial distribution
of charge.  The resulting emitted radiation spectrum is the product of the
Fourier transforms of these two functions:
\begin{equation}
\label{power}
P(\omega) \propto \left| E_{tot,\omega} \right|^2 \propto
\left|E_{point,\omega}\right|^2 \left|\lambda_\omega\right|^2,
\end{equation}
where $\vert E_{point,\omega} \vert$ is the field produced by an
accelerated point charge and $\lambda_\omega$ is the Fourier transform of
the charge density $\lambda(t)$ at the point on the particles' trajectory
where they are most closely directed towards the observer \cite{K18b}.  The
first factor on the right hand side of Eq.~\ref{power} is the smooth
spectrum shown in Fig.~\ref{curve} and the second factor gives the observed
structure on scales of tens to hundreds of MHz.  The fine spectral structure
produced by scintillation is a propagation effect, and is not included in
Eq.~\ref{power}.  For a point charge $\lambda(t)$ is a Dirac
$\delta$-function and $\lambda_\omega$ is a constant, so the spectrum of
Fig.~\ref{curve} is recovered.  For a uniform continuous distribution of
charge $\lambda(t) = \text{Constant}$ and $\lambda_\omega = 0$ for
$\omega \ne 0$ and there is no radiation.
\section{Open Questions}
\label{OQ}
Many observational facts about FRBs are known---brightness, spectra, pulse
shapes and widths, polarization, astronomical coordinates, dispersion and
(for a few FRBs) rotation measures.  Only FRB 121102 has been observed to
repeat; it is identified with a persistent radio source and a dwarf galaxy
at a cosmological redshift $z = 0.193$.  The distribution of FRBs on the sky
shows no obvious concentration (Fig.~\ref{bhandari}) and, allowing for
variations in search coverage, is likely consistent with isotropy.

But no one would say, beyond that basic phenomenology, that we understand
FRBs.  We should first ask what we want of a ``model''.  What should it
explain, and how much is it permitted to assume, or ignore?

The minimal requirement is that a model not be demonstrably wrong, either
observationally or theoretically.  It must at least be consistent with the
observed fluxes, durations, event rates, coordinates, spectra and
polarizations of FRBs.  It must also be consistent with the laws of physics,
both fundamental and phenomenological, as we have come to understand them.
It must invoke only the known or plausibly assumed components of the
universe.  Those are minimal requirements---it must not be inconsistent with
them.  Almost all published models pass this test (it's hard to publish
those that fail), but they are only minimal.

We want more than that.  Ideally, a model would have predicted the existence
of FRBs (none did, and prediction is rare in astrophysics, perhaps because
of our ignorance of initial conditions and the presence of turbulence).  It
should also make testable predictions.  That succeeded, once: The prediction
(based on isotropy on the sky and large dispersion measures) that FRBs
originate in the distant universe was verified by the identification of FRB
121102.  Many other predictions remain untested because they are beyond the
capabilities of existing instruments or because the arguments behind them
are insufficiently specific or quantitative.

A model should explain, perhaps by analogy, why FRBs result from the assumed
circumstances; it should be more than a ``Just so story''.  Evaluation of
plausibility is necessarily subjective.  Rather than advocate our favorite
models, we should ask questions of them.  This is partly to find testable
predictions so that the thicket of models can be thinned, but also because
asking questions, even if they are not answered, leads to insight and better
models.

Many questions, observational and theoretical, remain unanswered:
\begin{description}
	\item[Collimation] Is FRB emission collimated?  If so, it would
		relax the extreme energetic demands placed on pulsar giant
		pulse models.  In SGR models, it might explain the failure
		to observe a FRB during the December 27, 2004 outburst of
		SGR 1806-20, but see discussion in Sec.~\ref{SGR}.
		Collimation would be a convenient assumption, but would not
		help discriminate between those two classes of models, and
		would be hard to test.
	\item[Rotation Measures] What does the large RM of the repeating FRB
		121102 tell us about its environment?  Is it related to its
		repetition?
	\item[Dispersion Measures] Is the near-source contribution to FRB DM
		large enough to invalidate the inference (based on
		attributing most of the dispersion to the intergalactic
		medium) of distance from DM?
	\item[Is the Dispersion Measure of the Repeater Changing?] For one
		value of DM, presumably the correct value, the flux
		densities of many FRBs briefly drop to zero across the
		spectrum, appearing as vertical stripes in ``waterfall''
		plots like Fig.~\ref{gajjarf1} and (for FRB 170827)
		Fig.~\ref{170827}.  If such zero-power intervals are a real
		property of FRBs, then their DM can be determined precisely
		by requiring that these narrow stripes be accurately
		vertical.  Some models of FRB 121102 (for example, that it
		is a young neutron star surrounded by an expanding supernova
		remnant) predict changing DM, and can be constrained or
		verified by precise measures of DM at different epochs.  The
		DM of FRB 170827 has been measured to a precision of $\pm
		0.04$ pc cm$^{-3}$ \cite{F18}, roughly 100 times better than
		the uncertainties of the DM quoted for most FRB 121102
		bursts; can similar precision be obtained for FRB 121102?
	\item[Spectral Structure] Scintillation explains the higher
		resolution ($< 10$ MHz) spectral structure of FRBs.  Can it
		explain the 10--300 MHz structure, or is that intrinsic to
		the emission mechanism?
	\item[Spatial Distribution] Is the distribution of FRBs in space
		statistically homogeneous (aside from the effect of cosmic
		evolution at $z \gtrsim {\cal O}(1)$)?
	\item[Why does this FRB Differ from all other FRBs?] Why does FRB
		121102 repeat while no other FRB has been observed to
		repeat?  Does FRB 121102 manifest different temporal,
		spectral or polarization behavior?  If not, what explains
		their quantitative (repetition rate) difference?  Are they
		produced by different objects or mechanisms?
		Are they qualitatively or only quantitatively different?
	\item[Sources] Are FRBs produced by neutron stars?  If so, can they
		be also radio pulsars, soft gamma repeaters, or some other
		kind of neutron stars?  If not, what are their sources?
	\item[Applications] Can we learn about other fields of astrophysics
		from FRBs \cite{Mac15}?  Do they tell us anything about
		neutron star parameters at birth (rotation rate and field),
		the structure of interstellar and intergalactic plasmas,
		conditions near galactic nuclei, the distribution of matter
		in the Universe, cosmology, plasma turbulence and
		pulsars$\ldots$?
\end{description}
\section*{Acknowledgments}
I thank S. Bhandari, W. Farah, V. Gajjar, C. J. Law and E. Petroff for
permission to reproduce figures and D. Lorimer for valuable discussions.
%\begin{figure}[tb]
%%\epsfysize=9.0cm  THIS LINE COMMENTED OUT IN TEMPLATE
%\begin{center}
%\begin{minipage}[t]{8 cm}
%\epsfig{file=emblem.ps,scale=0.5}
%\end{minipage}
%\begin{minipage}[t]{16.5 cm}
%\caption{Cartoon of a nucleus, displaying the size of the nucleons as compared
%to the typical distance to nearest neighbors. Also indicated are the internal
%structure of nucleons and mesons.\label{fig1}}
%\end{minipage}
%\end{center}
%\end{figure}

\end{document}